\documentclass[11pt]{article}
\usepackage{amssymb}
\usepackage{graphicx}
\usepackage{layout}
\usepackage{bm}
\begin{document}

\title{The Sagnac Effect in curved space-times from an analogy with the Aharonov-Bohm Effect}
\author{Matteo Luca Ruggiero$^{1,2}$
\\
\small
$^1$ Dipartimento di Fisica, Politecnico di Torino,\\
\small $^2$ INFN, Sezione di Torino\\
\small e-mail: matteo.ruggiero@polito.it} \maketitle

\begin{abstract}
In the context of the natural splitting, the standard relative
dynamics can be expressed in terms of  gravito-electromagnetic
fields, which allow to formally introduce a gravito-magnetic
Aharonov-Bohm effect. We showed elsewhere that this formal analogy
can be used to derive the Sagnac effect in flat space-time as a
gravito-magnetic Aharonov-Bohm effect. Here, we generalize those
results to study the  General Relativistic corrections to the
Sagnac effect in some stationary and axially symmetric geometries,
such as the space-time around a weakly gravitating and rotating
source, Kerr space-time, G\"{odel} universe and Schwarzschild
space-time.
\end{abstract}

\small

\noindent Keywords: Sagnac Effect, Gravitoelectromagnetism,
Aharonov-Bohm
Effect, Gravito-Magnetic Aharonov-Bohm Effect.\\

\noindent PACS: 04.20.-q.

\normalsize

\section{Introduction}\label{sec:intro}

In previous papers \cite{rizzi03a,rizzi_ruggiero04} we showed that
the standard relative dynamics, in the background of the natural
splitting developed by Cattaneo \cite{cattaneo,catt1,catt2,
catt3,catt4}, can be described in terms of a
gravito-electromagnetic (GEM) formal analogy. In other words, the
time+space splitting allows to write the equation of motion of a
particle, in arbitrary space-times, in analogy with the equation
of motion of a charged particle acted upon by a "generalized"
Lorentz force, in flat space-time.

This analogy holds in full theory and, in linear approximation, it
corresponds to the well known analogy between the theory of
electromagnetism and the linearized theory of General Relativity
(GR) \cite{mashh1}. Indeed, gravito-electromagnetism in
relativistic gravity is somewhat ubiquitous, because of its "many
faces" \cite{jantzen92}, and various approaches
prove useful for studying gravitational phenomena in terms of
electromagnetic ones \cite{ruggiero02,tartaglia05}.

Our GEM approach to the study of relativistic dynamics leads to
what we called gravito-magnetic (GM) Aharonov-Bohm effect. Indeed,
in the past other authors
\cite{aharonov73,sakurai80,semon82,papini67,wisni67,stodolsky79,anandan77}
suggested the possibility of describing the phase shift and time
delay, induced by inertial and gravitational fields, in terms of
an analogy with the Aharonov-Bohm effect for the electromagnetic
field \cite{aharonovbohm59}. Our formalism applies to both
inertial and gravitational fields, holds in full theory and
generalizes the previous results, which were obtained, almost
always, at first order approximation (with respect to $c$, i.e. in
the slow motion approximation).

We already used this approach to obtain a derivation "by analogy"
of the Sagnac phase shift and time delay, for matter or light
beams counter-propagating on a round trip, in a rotating reference
frame in flat space-time \cite{rizzi03a,rizzi_ruggiero04}, and our
results are in agreement with the fully relativistic treatment of
the Sagnac effect given by Anandan \cite{anandan81}. In doing so,
we exploited the formal analogy between matter beams (or light
beams) counter propagating along circular trajectories in inertial
fields, and charged beams propagating in a region where a magnetic
vector potential is present. In other words, we showed that the
Sagnac effect may be thought of as a GM Aharonov-Bohm effect.

Here we are concerned with the generalization of those results to
curved space-times: we study the Sagnac effect in the  space-time
around a weakly gravitating and rotating source, in Kerr
space-time, in G\"{odel} universe and Schwarzschild space-time,
and we derive the Sagnac time-delay by following the analogy with
the Aharonov-Bohm effect.\footnote{Several descriptions of the Aharonov-Bohm effect have been developed (see for instance \cite{tourrenc77}); as for the Sagnac time-delay, we follow and extend the approach already considered in \cite{rizzi03b}.} In doing so, we recover some old results
\cite{cohen93,tartaglia98,kajari04}, and suggest that  our
approach can be applied  to arbitrary stationary and axially
symmetric geometries.

\section{Sagnac effect in  curved space-times}\label{sec:sagnac}
\begin{figure}[top]
\begin{center}
\includegraphics[width=6cm,height=5cm]{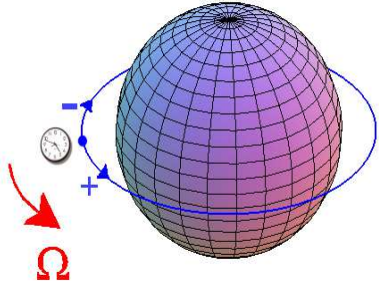}
\caption{\small We consider the  proper time difference (as
measured by a  clock rotating with constant angular velocity
$\Omega$) between the emission and absorption of the
co-propagating (+) and counter-propagating (-) beam.}
\label{fig:sphere}
\end{center}
\end{figure}
\normalsize

The Sagnac effect has been thoroughly studied in the past, and it
has been detected in many experiments (see \cite{rizzi_ruggiero04}
for a recent review). It is well known that when observing the
interference between light or matter beams (such as light beams,
electron or neutron beams and so on) counter-propagating in flat
space-time along a closed path in a rotating interferometer a
fringe shift\footnote{With respect to the interference pattern
when the device is at rest.} $\Delta \Phi$ arises. This phase
shift can be interpreted as a  time difference between the
propagation times (as measured by a clock at rest on the rotating
interferometer) of the co-rotating and counter-rotating beam.

By applying Cattaneo's splitting, the dynamics of  massive or
massless particles, relative to a given time-like congruence
$\Gamma$ of unit vectors $\bm{\gamma}(x)$, can be described in
terms of a "gravito-electromagnetic" fields
\cite{rizzi_ruggiero04}. By exploiting this fact, in
\cite{rizzi03a} we already showed that the Sagnac effect, for both
matter and light beams, counter-propagating on a round trip in a
interferometer rotating in flat space-time, may be obtained by
following a formal analogy with the Aharonov-Bohm effect. This
procedure can be generalized to study the Sagnac effect in curved
space-time, in order to obtain the GR corrections. Then, we can
study the interference process of matter or light beams in a
rotating frame in curved space-time in terms of the GM
Aharonov-Bohm effect. In other words,  this corresponds to
calculating the  GM Aharonov-Bohm phase shift (see
\cite{rizzi03a,rizzi_ruggiero04})
\begin{equation}
\Delta \Phi =\frac{2m\gamma _{0}}{c\hbar }\oint_{C}\widetilde{\bm{A}}_{G}\cdot {\rm d}%
\widetilde{\bm{x}}=\frac{2m\gamma _{0}}{c\hbar
}\int_{S}\bm{\widetilde{B}}_{G}\cdot {\rm d}\bm{\widetilde{S}},
\label{eq:gab}
\end{equation}
detected by a uniformly rotating interferometer, and the time
difference
\begin{equation}
\Delta T = \frac{2\gamma _{0}}{c^3}\oint_{C}\widetilde{\bm{A}}_{G}\cdot {\rm d}%
\widetilde{\bm{x}}=\frac{2\gamma
_{0}}{c^3}\int_{S}\bm{\widetilde{B}}_{G}\cdot {\rm
d}\bm{\widetilde{S}}, \label{eq:deltatau1}
\end{equation}
measured by a comoving observer provided with a standard
clock.

The time difference (\ref{eq:deltatau1}) corresponds to the Sagnac
time-delay: we notice that its expression clearly evidences the
"universal" character of the  Sagnac effect, since the mass $m$,
or more correctly, the energy\footnote{In (\ref{eq:gab}) $m$ is
the standard relative mass or standard relative energy of the
particles of the beams (see \cite{rizzi_ruggiero04}).} of the
particles of the interfering beams does not appear. So, the Sagnac
effect turns out to be an effect of the geometry of space-time,
and it can be considered universal, in the sense that it is the
same, independently of the physical nature of the interfering
beams.

As we said, eq. (\ref{eq:deltatau1}) expresses  the delay, for a
round trip, between the propagation time of the co-propagating and
counter-propagating beam (see Figure \ref{fig:sphere}). In what
follows, the two beams are supposed to have the same velocity (in
absolute value) with respect to the rotating frame. Furthermore,
the particles of the beams are supposed to be spinless, or their
spins are neglected.\footnote{Indeed, things are different when a
particle with spin, moving in a rotating frame or around a
rotating mass, is considered. In this case a coupling between the
spin and the angular velocity of the frame or the angular momentum
of the rotating mass appears (this effect is evaluated by Hehl-Ni
\cite{hehl90}, Mashhoon \cite{mashhoon88} and Papini
\cite{papinilibro}). Hence, the formal analogy leading to the
formulation of the GM Aharonov-Bohm effect holds only when the
spin-rotation coupling is neglected.}

It is easy to recognize that this analogy between the Sagnac
effect and the Aharonov-Bohm effect holds in arbitrary stationary
and axially symmetric geometries. Hence, in these cases, for
studying the Sagnac effect   it is sufficient to express the
space-time metric in coordinates adapted to a congruence of
rotating observers. Generally speaking, the metrics we deal with
are given in coordinates adapted to a congruence of asymptotically
inertial observers; if the coordinates are spherical, the passage
to a congruence of observers uniformly rotating in the equatorial
plane is obtained by applying the (azimuthal) coordinates
transformation
\begin{equation} \left\{
\begin{array}{rcl}
ct & = & ct', \\
r & = & r', \\
\varphi & = & \varphi' -\Omega t', \\
\vartheta & = & \vartheta',
\end{array}
\right.   \label{eq:sagnac_trasf_cord}
\end{equation}
where $\Omega$ is the (constant) angular velocity.

Then, it is simple to apply the formalism that we have already
used in \cite{rizzi03a,rizzi_ruggiero04}: the following steps give
a prescription for calculating the Sagnac effect in arbitrary
stationary and axially symmetric geometries in both flat and
curved space-times.\footnote{Greek indices run from 0 to 3, Latin
indices run from 1 to 3, the signature of the space-time metric is
$(-1,1,1,1)$.}
\begin{itemize}
\item Define the time-like congruence $\Gamma$ of rotating observers.\\

\item Express the space-time metric in coordinates adapted to
congruence of rotating observers.

\item Calculate the unit vectors field $\bm{\gamma}(x)$\\
\begin{displaymath}
\left\{
\begin{array}{l}
\gamma ^{{0}}\doteq\frac{1}{\sqrt{-g_{{00}}}}, \\
\gamma ^{i}\doteq0,
\end{array}
\right. \;\;\;\;\;\;\;\;\;\;\;\;\left\{
\begin{array}{l}
\gamma _{0}\doteq-\sqrt{-g_{{00}}}, \\
\gamma _{i}\doteq g_{i{0}}\gamma ^{{0}}.
\end{array}
\right.
\end{displaymath}

\item Calculate the gravito-magnetic vector potential

\begin{displaymath}
\widetilde{A}_{i}^{G}\doteq c^{2}\frac{\gamma _{i}}{\gamma _{0}}.
\end{displaymath}
\item Calculate the Sagnac time-delay as a GM Aharonov-Bohm effect
\begin{displaymath}
\Delta T = \frac{2\gamma _{0}}{c^3}\oint_{C}\widetilde{\bm{A}}_{G}\cdot {\rm d}%
\widetilde{\bm{x}}=\frac{2\gamma
_{0}}{c^3}\int_{S}\bm{\widetilde{B}}_{G}\cdot {\rm
d}\bm{\widetilde{S}}. \label{eq:deltatau11}
\end{displaymath}
\end{itemize}

We want to point out that both (\ref{eq:gab}) and
(\ref{eq:deltatau1}) have been obtained taking into account the
hypothesis that the two beams  have, locally,  the same relative
velocity\footnote{Generally speaking, we deal with wave packets,
hence the velocities we refer to, here and henceforth, are the
group velocities of these wave packets (see
\cite{rizzi_ruggiero04}, Sec. 3).}  (in absolute value) in  the
reference frame defined by the congruence $\Gamma$. In particular,
if we refer to a rotating frame in flat space-time, this
hypothesis coincides with the condition of "equal velocities in
opposite directions" that we imposed in order to obtain the Sagnac
effect by using a direct approach \cite{rizzi03b}.

In the following,  we calculate the Sagnac effect in some curved
space-times of physical interest (weak field space-time, Kerr
space-time,  G\"{o}del's universe, Schwarzschild space-time) by
applying the formalism described above. The calculations are
performed in full details for the case of the space-time around a
weakly gravitating and rotating source, to give an example.

Before going on, let us point out that, when dealing with weak
field, Schwarzschild and Kerr space-times, we consider beams
counter propagating around the source of gravitational field,  in
its equatorial plane, along circular space trajectories defined by
$r=R=constant$. In the case of G\"{o}del's universe, there is not
a localized source, but matter is everywhere. Nonetheless, because
of the spatial homogeneity of G\"{o}del solution we may consider,
again, without loss of generality, beams propagating along
circular trajectories $r=R=constant$ in the equatorial plane
defined by our choice of coordinates.

\subsection{The weak field around a rotating mass}\label{ssec:sagnac_curved_wf}

The space-time around a weakly  gravitating object of mass $M$ and
angular momentum $\bm{J}$ is given\footnote{See for instance
\cite{MTW},  Sec. 18.1, and also \cite{ruggiero02}.}
by\footnote{Here and henceforth, we use geometric units such that
$G=c=1$. Nevertheless, we keep $c$ and $G$ in the final results,
for the sake of clarity. }
\begin{eqnarray}
ds^2 & = & -\left(1-\frac{2M}{r'}
\right)dt'^2+\left(1+\frac{2M}{r'} \right)\left[dr'^2+r'^2
\left(d\vartheta'^2+\sin^2 \vartheta' d\varphi'^2 \right)\right] +
\nonumber \\ & & -\frac{4J}{r'}\sin^2 \vartheta' d\varphi' dt',
\label{eq:wf1}
\end{eqnarray}
where the spherical coordinates $(t',r',\vartheta',\varphi')$
(adapted to a congruence of a asymptotically inertial observers)
have been arranged in such a way that the angular momentum is
orthogonal to the equatorial plane.

The last term in (\ref{eq:wf1}) is responsible for the rotation of the local inertial frames,
the so called Lense-Thirring effect (see \cite{lensthir3}).

If we apply the transformation (\ref{eq:sagnac_trasf_cord}) to the
metric (\ref{eq:wf1}), after setting $\vartheta=~\frac{\pi}{2}$,
we obtain
\begin{eqnarray}
ds^2 & = & -\left[1-\frac{2M}{r}-\left(1+\frac{2M}{r}
\right)\Omega^2 r^2+\frac{4J\Omega}{r} \right]dt^2+ \nonumber \\
& & + \left(1+\frac{2M}{r} \right)\left(dr^2+r^2
d\varphi^2\right)+ 2\left[\left(1+\frac{2M}{r} \right)\Omega
r^2-\frac{2J}{r} \right] d\varphi dt. \nonumber \\ & &
\label{eq:wf2}
\end{eqnarray}
On evaluating the  components of the vector field
$\bm{\gamma}(x)$, along the trajectories $r=R=constant$, we have
\begin{eqnarray}
\gamma^0 & = & \gamma_J, \nonumber \\
\gamma_0 & = & -(\gamma_J)^{-1}, \nonumber \\
\gamma_\varphi & = & \left[\left(1+\frac{2M}{R}\right)\Omega
R^2-\frac{2J}{R}  \right]\gamma_J, \label{eq:gamma_weak1}
\end{eqnarray}
where
\begin{equation}
\gamma_{J} \doteq \left[1-\frac{2M}{R}-\left(1+\frac{2M}{R}
\right)\Omega^2 R^2+\frac{4J\Omega}{R} \right]^{-1/2},
\label{eq:gammaj2}
\end{equation}
and the corresponding component of the gravito-magnetic vector
potential is
\begin{equation}
\widetilde{A}^G_\varphi  =  -\left[\left(1+\frac{2M}{R}
\right)\Omega R^2-\frac{2J}{R} \right] \gamma^2_{J}.
\label{eq:agj1}
\end{equation}

On using physical units,  the phase shift (\ref{eq:gab}) turns out
to be
\begin{equation}
\Delta \Phi = \frac{4\pi m}{\hbar} \left[\left(1+\frac{2GM}{c^2R}
\right)\Omega R^2-\frac{2GJ}{c^2R} \right] \gamma_{J},
\label{eq:wf4}
\end{equation}
and the  time delay (\ref{eq:deltatau1}) is
\begin{equation}
\Delta T  = \frac{4\pi}{c^2}\left[\left(1+\frac{2GM}{c^2R}
\right)\Omega R^2-\frac{2GJ}{c^2R} \right] \gamma_{J},
\label{eq:wf5}
\end{equation}
or, by explicitly writing $\gamma_J$
\begin{equation}
\Delta T =\frac{4\pi}{c^2}\frac{\left[\left(1+\frac{2GM}{c^2R}
\right)\Omega R^2-\frac{2GJ}{c^2R}
\right]}{\left[1-\frac{2GM}{c^2R}-\left(1+\frac{2GM}{c^2R}
\right)\frac{\Omega^2 R^2}{c^2}+\frac{4GJ\Omega}{c^4R}
\right]^{1/2}}. \label{eq:wf6}
\end{equation}

In (\ref{eq:wf4}) and (\ref{eq:wf5}) we can distinguish two
contributions: the first one is proportional to the angular
velocity $\Omega$ of the observer, and the other one depends on
the absolute value of the angular momentum of the source $J$.

We see that even when $\Omega=0$ a time difference appears: this
is due to the rotation of the source of the gravitational field.
In other words
\begin{equation}
\Delta T_J \doteq \frac{4\pi}{c^2}\frac{\left( -\frac{2GJ}{c^2R}
\right)}{\left(1-\frac{2GM}{c^2R}\right)^{1/2}}
\label{eq:DeltaT_J1}
\end{equation}
is what a non rotating observer would obtain when measuring the
propagation time for a complete round trip of the two beams,
moving in opposite direction along circular orbits $r=R=constant$.
This time difference corresponds  to the so called
\textit{gravito-magnetic time delay}, which has been obtained by
Stodolski \cite{stodolsky79}, Cohen-Mashhoon \cite{cohen93} in
weak field approximation, and by Tartaglia \cite{tartaglia98}, by
a first order approximation of the time delay in Kerr space-time.

By inspection of the time delay (\ref{eq:wf6}) we recognize that
there is a critical angular velocity $\bar{\Omega}$ such that
$\Delta T=0$:
\begin{equation}
\bar{\Omega}=\frac{1}{R^2}\frac{\frac{2GJ}{c^2R}}{1+\frac{2GM}{c^2R}}
\label{eq:barOmega1}.
\end{equation}
$\bar{\Omega}$  corresponds to the angular velocity of the
"Locally Non Rotating Observers"\footnote{See, for instance,
\cite{MTW}, page 895.}
\begin{equation}
\Omega_{LNRO} \doteq -\frac{g_{t\varphi}}{g_{\varphi\varphi}}
\label{eq:OmegaLNRO1}
\end{equation}
of the weak field described by the metric (\ref{eq:wf1}).

These observers measure no time delay. A similar angular velocity
can be obtained also in Kerr space-time \cite{tartaglia98}.\\

\subsection{Kerr space-time}\label{ssec:sagnac_kerr}

Kerr solution \cite{kerr63} of Einstein equations describes the
space-time around a rotating black-hole or, more generally
speaking, around a rotating singularity. The classical form of
this solution is given in Boyer-Lindquist coordinates
\cite{boylin67} $(t',r',\vartheta',\varphi')$, which are adapted
to a congruence of asymptotically inertial observers
\begin{eqnarray}
ds^{2} &=&-(1-\frac{2M r'}{\rho ^{2}})dt'^{2}+\frac{\rho ^{2}}{\Delta }%
dr'^{2}+\rho ^{2}d\vartheta'^{2} + \nonumber \\ & & +\left(
r'^{2}+a^{2}+\frac{2M r'a^{2}\sin ^{2}\vartheta' }{\rho
^{2}}\right) \sin ^{2}\vartheta' d\varphi'^{2} -\frac{4M r'a\sin
^{2}\vartheta' }{\rho ^{2}}d\varphi' dt',  \nonumber \\ & &
\label{eq:kerr}
\end{eqnarray}
where  $\rho ^{2}=r'^{2}+a^{2}\cos ^{2}\vartheta' $, $\Delta
=r'^{2}-2M r'+a^{2}$, $a=J/Mc$,  $J$ is the absolute value of the
angular momentum, and the coordinates are arranged in such a way
that the angular momentum is perpendicular to the equatorial
plane.

On applying the prescription described above,  the phase shift
turns out to be
\begin{equation}
\Delta \Phi  = \frac{4\pi mc}{\hbar} \frac{\left[
-\frac{2GMa}{c^2R}+(R^2+a^2+\frac{2GMa^2}{c^2R})\frac{\Omega}{c}
\right]}{\left[1-\frac{2GM}{c^2R}+\frac{4GMa}{c^2R}\frac{\Omega}{c}-(R^2+a^2+\frac{2GMa^2}
{c^2R})\frac{\Omega^2}{c^2} \right]^{1/2}} ,
\label{eq:Deltaphi_kerr2}
\end{equation}
and the corresponding time delay is
\begin{equation}
\Delta T=\frac{4\pi}{c}\frac{\left[
-\frac{2GMa}{c^2R}+(R^2+a^2+\frac{2GMa^2}{c^2R})\frac{\Omega}{c}
\right]}{\left[1-\frac{2GM}{c^2R}+\frac{4GMa}{c^2R}\frac{\Omega}{c}-(R^2+a^2+\frac{2GMa^2}
{c^2R})\frac{\Omega^2}{c^2} \right]^{1/2}}.
\label{eq:DeltaT_kerr2}
\end{equation}

The time delay (\ref{eq:DeltaT_kerr2}) is in agreement with the
results obtained by Tartaglia \cite{tartaglia98}, who studied in
full details the GR corrections to the Sagnac effect in Kerr
space-time.

\subsection{G\"{o}del's universe}\label{ssec:godel}

G\"{o}del's solution  of Einstein's field equation describes,
roughly speaking, a rotating universe \cite{godel49}. This
solution, indeed, is not satisfactory on the observational
viewpoint, since it does not  agree with the expansion of the
universe and with  the cosmological observations of the cosmic
background radiation \cite{bunn96}. Another peculiarity of
G\"{o}del's solutions is that it admits closed time-like curves
 \cite{hawkingellis73}. Despite these pathologies, recently there
has been some interest about this exact solution of Einstein
equations \cite{kajari04}. As for us, since this solution
describes a rotating universe, it proves useful for studying the
corresponding gravito-magnetic features and, among them, the
Sagnac effect.

The metric of the G\"{o}del's solution can be written in the form
\cite{kajari04}
\begin{equation}
ds^2=-dt'^2+\frac{dr'^2}{1+\left(\frac{r'}{2a}
\right)^2}+r'^2\left[1-\left(\frac{r'}{2a} \right)^2 \right]
d\varphi'^2+dz'^2-2r'^2\frac{1}{\sqrt{2}a}dt' d\varphi',
\label{eq:godel1}
\end{equation}
where cylindrical coordinates $(t',r',\varphi',z')$ have been used
and $a>0$ is a constant having  the unit of a length. In our
analysis, we confine ourselves to the cylinder $r'<2a$, in order
to avoid problems with the chrono-geometric structure of the
G\"{odel}'s metric.

Then, following the approach described before, the phase shift
becomes
\begin{equation}
\Delta \Phi = \frac{4\pi m}{\hbar}
\frac{\left\{\Omega_G+\left[1-\left( \frac{R}{2a}\right)^2
\right]\Omega \right\}R^2}{\left\{1-\left[1-\left(
\frac{R}{2a}\right)^2 \right]\frac{R^2\Omega^2}{c^2} -2\frac{R^2
\Omega \Omega_G}{c^2}\right\}^{1/2}}, \label{eq:godel21}
\end{equation}
where
\begin{equation}
\Omega_G \doteq -\frac{c}{\sqrt{2}a}. \label{eq:omegaG11}
\end{equation}
The corresponding time delay turns out to be
\begin{equation}
\Delta T = \frac{4\pi}{c^2}\frac{\left\{\Omega_G+\left[1-\left(
\frac{R}{2a}\right)^2 \right]\Omega
\right\}R^2}{\left\{1-\left[1-\left( \frac{R}{2a}\right)^2
\right]\frac{R^2\Omega^2}{c^2} -2\frac{R^2 \Omega
\Omega_G}{c^2}\right\}^{1/2}}. \label{eq:godel31}
\end{equation}
The time delay (\ref{eq:godel31}) is in agreement with the results
given  by Kajari \textit{et al} \cite{kajari04}.\footnote{Notice
that their definition of $\Omega_G$ has a different sign with
respect to ours.}

Also in this case, we may obtain the velocity of rotation of the
observers that measure no time delay, i.e. the "Locally Non
Rotating Observers". In agreement with (\ref{eq:OmegaLNRO1}), from
(\ref{eq:godel31}), we get
\begin{equation}
\bar{\Omega} =-\frac{\Omega_G}{1-\left( \frac{R}{2a}\right)^2 }.
\label{eq:OmegaLNROgodel}
\end{equation}

The previous results, in practice, refer to the physical situation
in which the observer (or the detector) uniformly rotates relative
to the cosmological fluid, with angular velocity $\Omega$. So, if
we want to consider an observer at rest with respect to the
cosmological fluid of the G\"{o}del universe, it is sufficient to
set $\Omega=0$ in the previous formulae. Consequently  the phase
shift turns out to be
\begin{equation}
\Delta \Phi  = \frac{4\pi m}{\hbar} \Omega_G R^2,
\label{eq:godel2}
\end{equation}
and the corresponding time delay is
\begin{equation}
\Delta T  = \frac{4\pi}{c^2} \Omega_G R^2. \label{eq:godel3}
\end{equation}
Also the time delay (\ref{eq:godel3}) is in agreement with the
results by Kajari \textit{et al.} \cite{kajari04}.

In these calculations we have formally introduced the quantity
$\Omega_G$, whose dimensions are those of an angular velocity. It
is possible to show that $\Omega_G$ is the only non null
components of the vortex vector (see \cite{rizzi_ruggiero04}) of
the congruence of observers at rest with respect to the
cosmological fluid.  The vortex vector gives the (local) rate of
rotation\footnote{With respect to the proper time at any particle
of the reference fluid.} of the set of the neighbouring particles
of the congruence relative to the (local) compass of inertia. In
other words, the (constant) angular velocity $\Omega_G$ expresses
the rotation of the particles of the reference fluid with respect
to local gyroscopes. This means that, in G\"{o}del's universe,
matter rotates with constant angular velocity in the local
inertial frames of its comoving observers. The time delay
(\ref{eq:godel3}) is an effect of this rotation.

\subsection{Schwarzschild space-time}\label{ssec:schw}

The standard form of the classical Schwarzschild solution,
describing the vacuum space-time around a spherically symmetric
mass distribution\footnote{See, for instance, \cite{MTW},
Sec.23.6.} is
\begin{equation}
ds^2=-\left(1-\frac{2M}{r'} \right)dt'^2+\left(1-\frac{2M}{r'}
\right)^{-1}dr'^2+r'^2\left(d\vartheta'^2+\sin^2 \theta'
d\varphi'^2 \right), \label{eq:sch1}
\end{equation}
where $M$ is the mass of the source, and the coordinate
$(t',r',\vartheta',\varphi')$ are adapted to a congruence of
asymptotically inertial observers. In this case, our approach to
the study of the Sagnac effect leads to the phase shift
\begin{equation}
\Delta \Phi  =  \frac{4\pi m}{\hbar} \frac{\Omega
R^2}{\left(1-\frac{2GM}{Rc^2}-\frac{\Omega^2 R^2}{c^2}
\right)^{1/2}}, \label{eq:sagnacsch1}
\end{equation}
and the corresponding time delay turns out to be
\begin{equation}
\Delta T= \frac{4\pi }{c^2} \frac{\Omega
R^2}{\left(1-\frac{2GM}{Rc^2}-\frac{\Omega^2 R^2}{c^2}
\right)^{1/2}}. \label{eq:sagnacsch2}
\end{equation}
We see that if $\Omega=0$, i.e. if we measure the propagation time
in a non rotating frame, no Sagnac effect arises. In other words,
the propagation is symmetrical in both directions.

\section{Conclusions}\label{sec:conclusions}

In the context of natural splitting, the relative formulation of
dynamics can be expressed in terms of GEM fields; this analogy,
which holds in full theory, leads to the formulation of the GM
Aharonov-Bohm effect. We have exploited this fact to give a
derivation of the Sagnac effect based on this  formal analogy, in
stationary and axially symmetric geometries. In doing so, we have
showed that the Sagnac effect can be interpreted as a GM
Aharonov-Bohm effect in both flat and curved space-time.

In particular, we have studied the GR corrections to the Sagnac
effect in  the space-time around a weakly gravitating and rotating
source, in Kerr space-time,  in G\"{odel} universe and in
Schwarzschild space-time.

Our results are in agreement with those available in the
literature and generalize some old approaches, where the analogy
with the Aharonov-Bohm effect was limited to a first order
approximation.

The Sagnac effect has a "universal" character: in other words, the
Sagnac time-delay is the same, independently of the physical
nature and velocities of  the interfering beams, provided that the
latter are locally the same, in absolute value, as seen in the
rotating frame, as we explained in full details, by using a direct
approach, in \cite{rizzi03b}. In our formalism this universality
is expressed by eq. (\ref{eq:deltatau1}), where the mass (or the
energy) of the particles of the beams does not appear, and the
time delay is explained as an effect of the geometrical background
in which the beams propagate: in this fact we recognize the close
analogy with the Aharonov-Bohm effect, which has a purely
geometrical interpretation \cite{tourrenc77}. Hence, the
geometrical approach clearly points
out the universal character of the Sagnac effect.\\

\textbf{Acknowledgments:} The author would like to thank two
anonymous referees for their suggestions which contributed to
improve the paper.

\end{document}